\documentclass[aps,prl,twocolumn,superscriptaddress]{revtex4-2}
\usepackage{amssymb}
\usepackage{suffix}
\usepackage{float}
\usepackage{mathtools}
\usepackage[utf8]{inputenc}
\usepackage{booktabs}
\usepackage{cases}
\usepackage[multiple]{footmisc}
\usepackage{dcolumn}
\usepackage{color,soul}
\usepackage{rotating}
\usepackage{perpage}
\usepackage{xcolor}
\usepackage{soul}
\usepackage{ulem}
\usepackage{tikz}
\usepackage[T1]{fontenc}
\usepackage{etoolbox}
\usepackage{graphics}
\usepackage{siunitx}
\usepackage{hyperref}
\usepackage{float}	
\usepackage{collref}
\usepackage{multirow}
\usepackage{mathtools}
\usepackage{bm}
\usepackage{url}

\usepackage{tikz}
\usepackage{tikz-3dplot}
\usepackage{changes}
\usepackage{physics}

\newcommand*\rfrac[2]{{}^{#1}\!/_{#2}}

\begin{document} 
	\title{Laser-enhanced magnetism in SmFeO$_3$}
	\author{Mohsen Yarmohammadi}
	\email{mohsen.yarmohammadi@utdallas.edu}
	\affiliation{Department of Physics, The University of Texas at Dallas, Richardson, Texas 75080, USA}
	\author{Marin Bukov}
	\email{mgbukov@pks.mpg.de}
	\affiliation{Max Planck Institute for the Physics of Complex Systems, N\"othnitzer Str.~38, 01187 Dresden, Germany}
	\author{Vadim Oganesyan}
	\email{vadim.oganesyan@csi.cuny.edu}
	\affiliation{Physics program and Initiative for the Theoretical Sciences, The Graduate Center, CUNY, New York, NY 10016, USA}
	\affiliation{Center for Computational Quantum Physics, Flatiron Institute, 162 5th Avenue, New York, NY 10010, USA}
	\affiliation{Department of Physics and Astronomy, College of Staten Island, CUNY, Staten Island, NY 10314, USA}
	\author{Michael H. Kolodrubetz}
	\email{mkolodru@utdallas.edu}
	\affiliation{Department of Physics, The University of Texas at Dallas, Richardson, Texas 75080, USA}
	\date{\today}
	
	\begin{abstract}
		To coherently enhance inherent weak magnetic interactions in rare-earth orthoferrite SmFeO$_3$ as a functional material for spintronic applications, we simulate the dissipative spin dynamics that are linearly and quadratically coupled to laser-driven infrared-active phonons. When linear coupling dominates, we discover a magnetophononic dynamical first-order phase transition in the nonequilibrium steady state which can inhibit strong enhancement of magnetic interactions. By contrast, when quadratic spin-phonon coupling dominates, no phase transition exists at experimentally relevant parameters. By utilizing a chirp protocol, the phase transition can be engineered, enabling stronger magnetic interactions. We also discuss the route for experimental observation of our results.
	\end{abstract}
	
	\maketitle
	{\allowdisplaybreaks
            \textit{Introduction.}---In recent years, the prospect of using spins as alternative information carriers has led to the emerging field of spintronics, which promises next-generation digital processing units that are fast, robust, and energy efficient~\cite{ZHENG20031}. In solid materials exhibiting magnetic properties, the electronic spins interact with lattice vibrations~(phonons) opening up a backaction channel, through electron orbital hybridization. 
   
            The famous magnetic material class whose role in spintronics is as significant as silicon in electronics is the rare-earth orthoferrites~(perovskite oxides) RXO$_3$, with magnetic rare-earth ion R and a magnetic ion X~\cite{Li18}. A primary example is SmFeO$_3$, where at low temperatures an unusual strong nonlinear spin-phonon coupling~(SPC) arises from the cross-talk between the two magnetic iron and samarium ions~\cite{Weber2022}, in contrast to most examples that deal only with one magnetic lattice and linear SPC. 
            
            Even though the nonlinear SPC in SmFeO$_3$ is strong, the magnetic exchange interactions including Sm spins are weak~\cite{https://doi.org/10.48550/arxiv.2211.13528}. Enhancing these interactions will enhance the effects controlled by the unusual nonlinear SPC. This is necessary to leverage their unusually large spin interactions for the design of material properties from a scientific perspective, as they exhibit intriguing features such as the measurable magnetoelectric effect~\cite{Bousquet_2016,doi:10.1080/00150193.2017.1283171,Fiebig2016,doi:10.1063/1.3502547,Spaldin2017}, and for applications in industry, e.g., in solar cells and spintronics~\cite{SURESHKUMAR2021940,Wu2021,Kimel2005}. 
            
            A conventional way is to apply pressure~\cite{Zayed2017} or magnetic field~\cite{Romhanyi2015}. In contrast to these static approaches, dynamic control techniques emerge as a new frontier to induce and observe a broader array of novel phenomena that only exist away from equilibrium~\cite{Afanasiev2021,PhysRevLett.127.077202,PhysRevB.93.174117}. Since spin motions in solids typically have a timescale that coincides with the THz spectrum region, detection and manipulation of spins by optical means is indispensable. 
		
            In this Letter, we propose a nonequilibrium magnetophononic mechanism to manipulate inherent weak spin-spin interactions in SmFeO$_3$ via laser-driven infrared-active~(IR-active) phonons that play the central role of a mediator for dynamic control of spins. We first show that magnetic interactions are enhanced in SmFeO$_3$ for the on-resonance laser-phonon coupling regime. More importantly, we discover a \textit{dynamical magnetophononic first-order phase transition} in SmFeO$_3$ for the off-resonance laser-phonon coupling regime as a more efficient way compared to the on-resonance one to further enhance magnetism. We also discuss the origin of the transition from inherent non-linearities in the mechanism as well as a feasible solution in the experiment to observe these phenomena. Finally, we propose a linear chirp drive to further enhance the magnetic response of the system.

            \textit{Model.}---We consider SmFeO$_3$ at a very low temperature whose weak spin interaction between Sm$^{3+}$ and Fe$^{2+}$ ions can be well-described by an anisotropic antiferromagnetic~(AFM) Heisenberg $S = 5/2$ chain~(XXZ model)~\cite{Cao2014,Weber2022,https://doi.org/10.48550/arxiv.2211.13528,PhysRevLett.130.106902}. Vibrational modes are excited at an initial time by a THz continuous wave laser that creates a coherent phonon field~\cite{PhysRev.131.2766,yarmohammadi2020dynamical,yarmohammadi2023nonequilibrium,PhysRevB.107.174415} on every site $\ell$, which couples both linearly and quadratically to the nearest isotropic~(in-plane) and anisotropic~(easy-axis) spin interactions. Hence, the total Hamiltonian that governs the dynamics comprises four terms\setlength{\abovedisplayskip}{3pt}	\setlength{\belowdisplayskip}{1pt}	\begin{equation}\label{eq_1}
			\mathcal{H} = \mathcal{H}_{\rm s}  + \mathcal{H}_{\rm ph} + \mathcal{H}_{\rm sp} + \mathcal{H}_{\rm lp}\, ,
		\end{equation} where $\mathcal{H}_{\rm s} = \sum_{\langle \ell,j \rangle} \left(J \left[S^x_\ell S^x_j + S^y_\ell S^y_j\right] + J \Delta S^z_\ell S^z_j\right)$ is the XXZ model with the spin operators $S^\alpha$ for any flavor $\alpha = \{x,y,z\}$; $J \simeq 1.25$ meV describes the in-plane isotropic exchange coupling, $\Delta > 1$ describes the anisotropy of the AFM interactions. 
        
        In the above Hamiltonian, $\mathcal{H}_{\rm ph} = \omega_0 \sum_\ell a^\dagger_\ell a_\ell$ characterizes a local Einstein IR-active phonon mode with frequency $\omega_0$ described by the boson creation operator $a^\dagger_\ell$. Unfortunately, no IR experimental spectra are available for SmFeO$_3$ to select a specific phonon frequency. However, it has been found that most of the IR-active modes in SmFeO$_3$ are located far above the spin band~\cite{https://doi.org/10.48550/arxiv.2211.13528}. Therefore, we choose $\omega_0 > 2 J \Delta$ in what follows: e.g., for $\Delta = 1.2$ throughout the paper, we set $\omega_0/J = 2.5$.
        
        Assuming the same symmetries as in the static spin Hamiltonian, there are two possible contributions to the spin-phonon coupling part to locally couple spins to the excited linear and quadratic lattice vibrations: $\mathcal{H}_{\rm sp}^{\rm iso} = \sum_{\langle \ell,j \rangle} \big[g_{\rm l} (a^\dagger_\ell +a_\ell)+g_{\rm q} (a^\dagger_\ell +a_\ell)^2\big] \left[S^x_\ell S^x_j + S^y_\ell S^y_j \right]$ and $\mathcal{H}_{\rm sp}^{\rm aniso} = \sum_{\langle \ell,j \rangle} \big[g_{\rm l} (a^\dagger_\ell +a_\ell)+g_{\rm q} (a^\dagger_\ell +a_\ell)^2\big] \left[ \Delta S^z_\ell  S^z_j \right]$; $g_{\rm l}$~($g_{\rm q}$) is the linear~(quadratic) SPC strength. Since we have no microscopic values for the different coupling strengths, we consider them separately to analyze their individual effects. Both quadratic SPC and spin-spin interaction terms are even under inversion symmetry. For a general geometry including a perpendicular atomic motion to the bond, inversion symmetry also includes a linear SPC to the spin-spin interaction. 
        
        Finally, the laser is linearly coupled to the phonon via $\mathcal{H}_{\rm lp} =  \sum_\ell \mathcal{E}(t) (a^\dagger_\ell + a_\ell)$, where $\mathcal{E}(t) = \mathcal{A}_0 \cos(\omega\,t)$ describes the pump laser with amplitude $\mathcal{A}_0$ and frequency $\omega$. Following the Lindemann criterion~\cite{Lindemann1910} to avoid lattice melting, phonon occupation should not reach a too large number. For this reason, we set $\mathcal{A}_0/J \ll 1$ in our simulations. 
        
        To balance energy transfer and avoid heating effects related to driving and coherence preservation, the entire system is coupled to a bath comprised of other, undriven phonons. More details of the model, including the reciprocal-space representation can be found in the Supplemental Material~(SM)~\cite{SM}.
        
        We employ the zeroth order expansion of the Holstein-Primakoff transformation~\cite{PhysRev.58.1098} within linear spin-wave theory to explore phononically driven magnons. Since the population of excited magnons is low at low temperatures, we neglect the magnon-magnon interaction. Due to the long wavelength of the pump field compared to the lattice scale, we assume a local dispersionless phonon mode~($a^\dagger_0$) coupled to the laser. On the other hand, since the expectation value of the phonon occupation is proportional to the number of sites ($L = 2501$ in our simulation), the relative quantum fluctuations, proportional to $1/\sqrt{L}$, tend to zero in the thermodynamic limit ($L \to \infty$). We therefore employ a mean-field approximation~\cite{yarmohammadi2020dynamical} to decouple the SPC term acting on the phonon and spin.
		
		We simulate the time-evolution of physical observables of the model for both the spin and phonon sectors, using the adjoint Lindblad quantum master equation with phenomenological damping rates $\gamma_{\ell} = \gamma_{\rm ph},\gamma_{\rm s}$ for an arbitrary observable $O(t)$: $\langle \dot{O}\rangle (t) =  {} i\langle[\mathcal{H},O(t)]\rangle  +  \frac{1}{2}\sum_{\ell} \gamma_{\ell} \Big<\big[\mathcal{L}_{\ell}^{\dagger},O(t)\big]\mathcal{L}_{\ell}  + \mathcal{L}_{\ell}^{\dagger}\big[O(t),\mathcal{L}_{\ell}\big]\Big>$, where $\mathcal{L}_{\ell} = \{a^\dagger_0,a_0\}$ and $\{\mathcal{B}^\dagger_k,\mathcal{B}_k\}$ are the phonon- and Bogoliubov magnon-nonconserving time-independent Lindblad jump operators that relax the system towards its ground state by employing $\{\mathcal{L}_{\ell} = {} a^\dagger_0$ and $a_0  \mapsto  \gamma_{\ell} = \gamma_{\rm ph}\, \mathcal{N}_0$ and $\gamma_{\rm ph} [1+\mathcal{N}_0]$\} and $\{\mathcal{L}_{\ell} = {} \mathcal{B}^\dagger_k$ and $\mathcal{B}_k \mapsto  \gamma_{\ell} = \gamma_{\rm e}\, \mathcal{N}_k$ and $\gamma_{\rm e} [1+\mathcal{N}_k]\}$, where $\mathcal{N}_0$ and $\mathcal{N}_k$ is the average number of phonon and magnon with energy $\omega_{0}$ and $\varepsilon_k = {} 2 J \sqrt{\Delta^2 - \cos^2k }$, respectively. Given the weakness of spin damping compared to direct phonon decay, we set $\gamma_{\rm s} < \gamma_{\rm ph}$. Moreover, due to the weak coupling of a single optical phonon to the ensemble of phonons~(bath), we set $0.01 < \gamma_{\rm ph}/\omega_0 < 0.05$. Last, to treat the properties of nonequilibrium steady states~(NESS), we take the average of the late-time signals over one period $T_p$ via $\overline{O} = T_p^{-1} \int^{t+T_p}_t O(t)\mathrm dt$. \begin{figure}[t]
		\centering
		\includegraphics[width=0.75\linewidth]{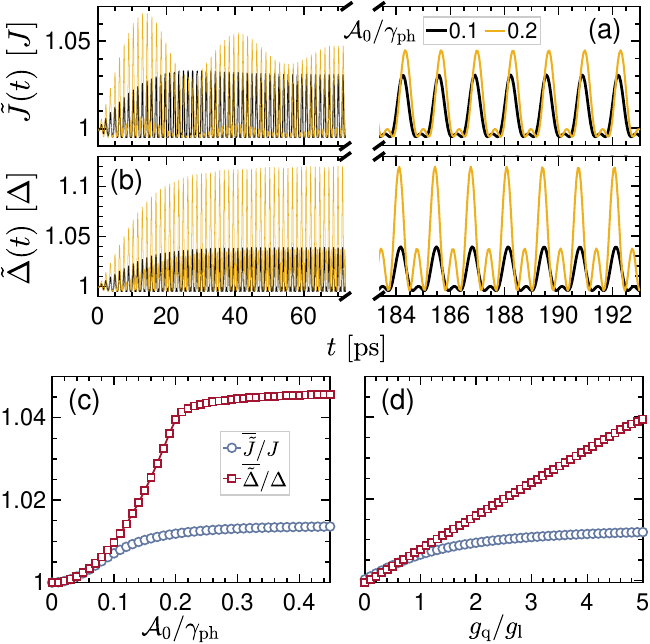}
		\setlength\abovecaptionskip{1pt}
		\caption{\textbf{Laser manipulation of magnetism in SmFeO$_3$ at low temperatures.} [(a),(b)] Laser amplitude dependency of the temporal isotropic~($J$) and anisotropic~($\Delta$) magnetic interactions with dominant quadratic SPC $g_{\rm q}/g_{\rm l} = 5$. Time averaging in the NESS gives interactions as a function of laser amplitude [(c), with $g_{\rm q}/g_{\rm l} = 5$] and SPC~[(d), with $\mathcal{A}_0/\gamma_{\rm ph} = 0.2$]. The inherent spin feedback effect in the magnetophononic mechanism leads to a saturated spin response to the laser field in the strong coupling regime, while a nonlinear~(linear) modulation of isotropic~(anisotropic) response to the SPCs appears due to different matrix elements of spin excitations, see Eq.~\eqref{eq_4}. Here, $\omega = \omega_0 = 2.5 J$, $\Delta = 1.2$, $\gamma_{\rm ph}/\omega_0 = 0.05$, and $\gamma_{\rm s}/J = 0.01$.
		} 
		\label{f1}
	\end{figure}\begin{figure*}[t]
		\centering
		\includegraphics[width=0.95\linewidth]{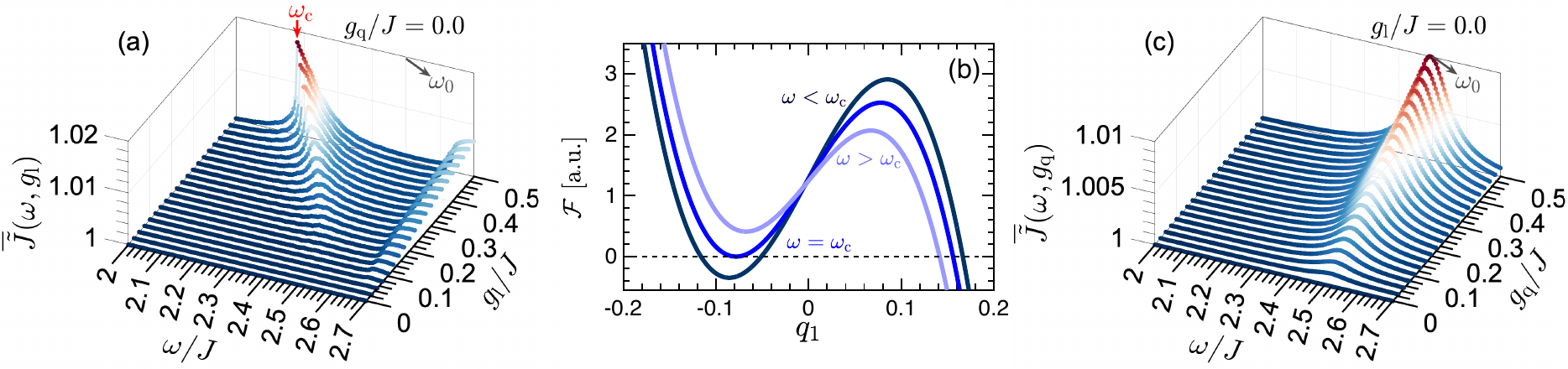}
		\setlength\abovecaptionskip{1pt}
		\caption{\textbf{Dynamical magnetophononic first-order phase transition.} 
			(a) Isotropic magnetic interaction as a function of drive frequency for various linear SPCs at $g_{\rm q}/J = 0.0$; at large $g_{\rm l}$, the appearance of the cusp signals the first-order phase transition. (b) Effective force in the NESS acting on the driven-dressed-damped phonon at $g_{\rm l}/J = 0.5$ and $g_{\rm q}/J = 0.0$ expanding magnons around the dominant $k=0$ mode, characterized by the first harmonic of phonon displacement $q_1$, to confirm the first-order phase transition at $\omega_{\rm c}$. (c) The same as (a) but vs.~quadratic SPCs at $g_{\rm l}/J = 0.0$ showing weak shifts of the phonon resonance leading to the absence of a phase transition. Here, $\omega_0 = 2.5 J$, $\Delta = 1.2$, $\mathcal{A}_0/\gamma_{\rm ph} = 0.1$, $\gamma_{\rm ph}/\omega_0 = 0.05$, and $\gamma_{\rm s}/J = 0.01$.
		} 
		\label{f2}
	\end{figure*}

	\textit{Results.}---Excitation of the IR-active phonons directly by the laser leads to a modulation of both magnetic coupling constants $J$ and $\Delta$, given by\setlength{\abovedisplayskip}{3pt}	\setlength{\belowdisplayskip}{1pt}	\begin{subequations}\label{eq_3}
		\begin{align}
			\tilde{J}(t)/J = {} &1 + g_{\rm l}q_{\rm ph,{\rm iso}}(t)  + g_{\rm q} q^{2}_{\rm ph,{\rm iso}}(t)\, ,\\
			\tilde{\Delta}(t)/\Delta = {} & 1 + g_{\rm l}q_{\rm ph,{\rm aniso}}(t)  + g_{\rm q} q^{2}_{\rm ph,{\rm aniso}}(t)\, ,
		\end{align}
	\end{subequations}where $q_{\rm ph}(t) = \langle L^{-1/2} (a^\dagger_0  + a_0) \rangle (t)$ describes the coherent lattice displacement~(oscillation) of the zero-momentum phonon mode~\cite{SM}. After solving the tightly coupled equations of motion between subsectors, as a result, magnons acquire dynamics via laser-driven phonons such that the properties of the spin band can be dynamically engineered. In Eq.~\eqref{eq_3}, the superscript ``iso'' (``aniso'') indicates that data is taken with SPC given by $\mathcal{H}_{\rm sp}^{\rm iso}$ ($\mathcal{H}_{\rm sp}^{\rm aniso}$). Note that, since isotropic SPC gives strong modification of $J$ while anisotropic SPC gives strong modification of only $\Delta$, all data for $\tilde J(t)$ ($\tilde \Delta(t)$) is taken using only $\mathcal{H}_{\rm sp}^{\rm iso}$ ($\mathcal{H}_{\rm sp}^{\rm aniso}$) throughout this paper.
	
	Figure~\ref{f1} demonstrates the modulation of $J$ and $\Delta$ in the course of the time evolution on short (left part of panels a and b) to long (right part of panels a and b) timescales, as the system approaches its long-time steady state at $t \simeq 100$ ps. We first consider the on-resonance laser-phonon coupling~(LPC) regime at $\omega = \omega_0$. Under the laser field, both exchange interactions reach a NESS, associated with coherent oscillations of about 5\% above the bare interaction strengths, as shown in Figs.~\ref{f1}(a,b). Coupling between spins and phonons already appears in the transient~(early-time) dynamics. The amplitude of oscillations increases with the laser amplitude in both interactions; the phase of oscillations in isotropic and anisotropic interactions slightly shifts to the right and left, respectively, due to a negative sign difference in the matrix elements of the SPC part, see Eq.~\eqref{eq_4}. The change in the occupation of phonons and then magnons~(following the modulated interactions) with the SPC is a direct consequence of hybridization effects. 
 
    Energy naturally flows from the laser to the spin system until a steady state is reached, modifying the effective exchange interactions. Since the laser fluence is proportional to $\mathcal{A}_0^2$, we would expect $J$ and $\Delta$ to scale as $\mathcal{A}_0^2$ too. Such a response is satisfied up to $\mathcal{A}_0/\gamma_{\rm ph} = 0.1$ and $0.2$, respectively, for $J$ and $\Delta$, but is saturated at stronger laser fields, as shown in Fig.~\ref{f1}(c). This can be understood from an inherent spin feedback effect in the strong coupling regime, where spin excitations shift the phonon resonance and inhibit further excitations. 
 
    A key point visible in the data is that quadratic SPC is significantly more effective at modifying magnetic interactions, both isotropic ($J$) and anisotropic ($\Delta$) when the phonon is driven resonantly, as shown in Fig.~\ref{f1}(d). This is partially due to linear SPC having a stronger effect on the phonon resonance frequency, as we will see, suggesting that perhaps we can modify interactions more strongly by driving off-resonance. 
    
    It is common to pump the system at variable driving frequencies $\omega$ in most ultrafast experiments. As we now show, that does not occur without challenges, and in fact, can lead to a dynamical phase transition in the NESS. We fix the phonon above the spin-band at $\omega_0/J = 2.5$ and scan the laser frequency to probe mutual dressing effects. For $g_{\rm q} = 0$, the laser electric field acting on the phonon rapidly becomes dressed with the feedback from the spin,\setlength{\abovedisplayskip}{3pt}	\setlength{\belowdisplayskip}{1pt}	\begin{equation}
		\widetilde{\mathcal{E}}(t) = {} \mathcal{E}(t) + \frac{g_{\rm l}}{L} \sum_k \big[ \mathcal{R}_k n_{{\rm s},k} (t) +  \mathcal{S}_k x_{{\rm s},k} (t)\big]\, ,
	\end{equation}where $n_{{\rm s},k}(t)$ is the magnon density, $x_{{\rm s},k}$ is the magnon pair creation observable, and \setlength{\abovedisplayskip}{3pt}	\setlength{\belowdisplayskip}{1pt}	\begin{subequations}\label{eq_4}
 \begin{align}
     \mathcal{R}^{\rm iso}_k = {} &-4 J \cos^2k/\varepsilon_k\, \, , \,\,
     \mathcal{R}^{\rm aniso}_k = {} 4 J \Delta^2 /\varepsilon_k\, ,\\
     \mathcal{S}^{\rm iso}_k = {} &2 J\Delta \cos k/\varepsilon_k\, \, , \,\,
     \mathcal{S}^{\rm aniso}_k = {} -2 J \Delta \cos k/\varepsilon_k\, ,
      \end{align}
 \end{subequations}are the matrix elements for the spin excitations. This implies that the interplay between phonon and laser reflects the tendency of laser energy to flow to the magnons; as a result, the resonance peak at $\omega  = \omega_0$ is shifted to lower frequencies due to the nonequilibrium magnon occupation, Fig.~\ref{f2}(a). 
 
 At strong linear SPCs, we observe a rapid jump of isotropic magnetic interaction $J$ at a critical drive frequency $\omega_{\rm c}$ -- doubling the size of its strength compared to on-resonance modulation -- accompanied by the creation of 2\% magnon density per site~(see Sec.~S3 of the SM). Such a jump corresponds to a dynamical first-order phase transition, similar to one that some of us found in previous work on a driven fermion chain coupled quadratically to lattice vibrations~\cite{yarmohammadi2023nonequilibrium}. In contrast to the dissipation-induced nonlinearity found in the fermion chain, there is a direct nonlinearity in the mean-field Hamiltonian, which leads to the phase transition in a model with only linear SPC. Therefore, unlike for the fermion chain, the mechanism that drives the phase transition reported here is inherent to the spin system, and the drive only serves as a trigger. The peak near $\omega/J \simeq 2.65$ is the fingerprint of the lower two-magnon band edge.

Next, we strive to understand the mechanism behind the appearance of the phase transition in the NESS, by evaluating the dominant harmonic of the driven-dressed-damped phonon. We derive an effective theoretical picture for the dynamics by approximating the magnons with their single, dominant, zero-momentum ($k = 0$) mode. To this end, we proceed with the Fourier decomposition of phonon displacement in the NESS, $q_{\rm ph}(t) = \overline{q_{\rm ph}} + \sum_{n \neq 0} q_n e^{i n \omega t}$, where $q_1$ is the dominant first harmonic. This leads to the following cubic equation\setlength{\abovedisplayskip}{3pt}	\setlength{\belowdisplayskip}{1pt}	\begin{equation}
\label{eq_5}
\mathcal{F} = a q^3_1 + b q^2_ 1 + c q_1 + d = 0\, ,
\end{equation} where the hybridized driving, coupling and damping processes are described by: \setlength{\abovedisplayskip}{3pt}	\setlength{\belowdisplayskip}{1pt}	\begin{eqnarray*}
a &=& 4 g^2_{\rm l} \mathcal{S}^2_{k=0} [\gamma_{\rm s}(\omega^2 - \omega^2_0 -\rfrac{\gamma^2_{\rm ph}}{4}) + \gamma_{\rm ph} \omega^2], \\
b &=&  -4 \mathcal{A}_0 \omega_0 \gamma_{\rm s} g^2_{\rm l}  \mathcal{S}^2_{k=0}, \\ 
c &=& \gamma_{\rm s} [\omega^2 - \omega^2_0 -\rfrac{\gamma^2_{\rm ph}}{4}] [\omega^2 - 4\varepsilon^2_{k=0} -\gamma^2_{\rm s}] \\
&& + \gamma_{\rm ph}\gamma^2_{\rm s} \omega^2 - 4 \gamma_{\rm s} g^2_{\rm l} \mathcal{S}^2_{k=0} \omega_0 \varepsilon_{k=0}, \\
d &=& - \mathcal{A}_0 \omega_0 \gamma_{\rm s} [\omega^2 - 4\varepsilon^2_{k=0} -\gamma^2_{\rm s}].
\end{eqnarray*} Equation~\eqref{eq_5} behaves like an effective force acting on the phonon in the driven-coupled-damped regime of the system. Thus, to sharply characterize the phase transition, we track the solution of this equation preceding that $\omega_{\rm c}$. Vanishing the effective force solutions at $\omega_{\rm c}$ is evidence for the phase transition, see Fig.~\ref{f2}(b). Not only does the effective force theory produce the origin of the transition, but it also provides an overall quantitative proxy for the exact $\omega_{\rm c}$.

It is important to note that near the first-order phase transition, we find a fast initial growth of dynamics of observables for $t \lessapprox 1$ ns due to enhanced magnon density and phonon occupation through linear SPC, see Sec.~S3 of the SM. This remarkable behavior implies that an effective hybridized state forms on transient timescales to anticipate the phase transition. The same analysis, applied to the coupling of the phonon to the anisotropic magnetic interaction, appears with a dynamical profile somewhat similar (albeit of opposite sign for the presence of quadratic SPC) to the isotropic one. 

Turning to the dynamical features of the system in the presence of quadratic SPC, in Fig.~\ref{f2}(c) we scan the system's response to the drive frequency at $g_{\rm l} = 0$. In contrast to the linear SPC, the drive pushes the system into a simple NESS with $g_{\rm q}$. Importantly, no phase transition occurs in this case, at least within the small to intermediate $g_{\rm q}$ regime. This weak dynamical response of the nonlinearly coupled spin-phonon system implies the existence of a quasi-decoupled phase in the system, which differs significantly from the nonanalytic response of the quadratic electron-phonon coupling in driven materials~\cite{yarmohammadi2023nonequilibrium,Sous2021,Kennes2017}. Thus, dominant quadratic SPC in SmFeO$_3$ should allow for the manipulation of the spin band in the on-resonance LPC regime while avoiding the phase transition in the off-resonance LPC regime. This, in turn, inhibits spin band engineering for cases where linear SPC dominates. Still, the presence of both linear and quadratic SPCs does not inhibit the appearance of the first-order phase transition.\begin{figure}[t]
\centering
\includegraphics[width=0.75\linewidth]{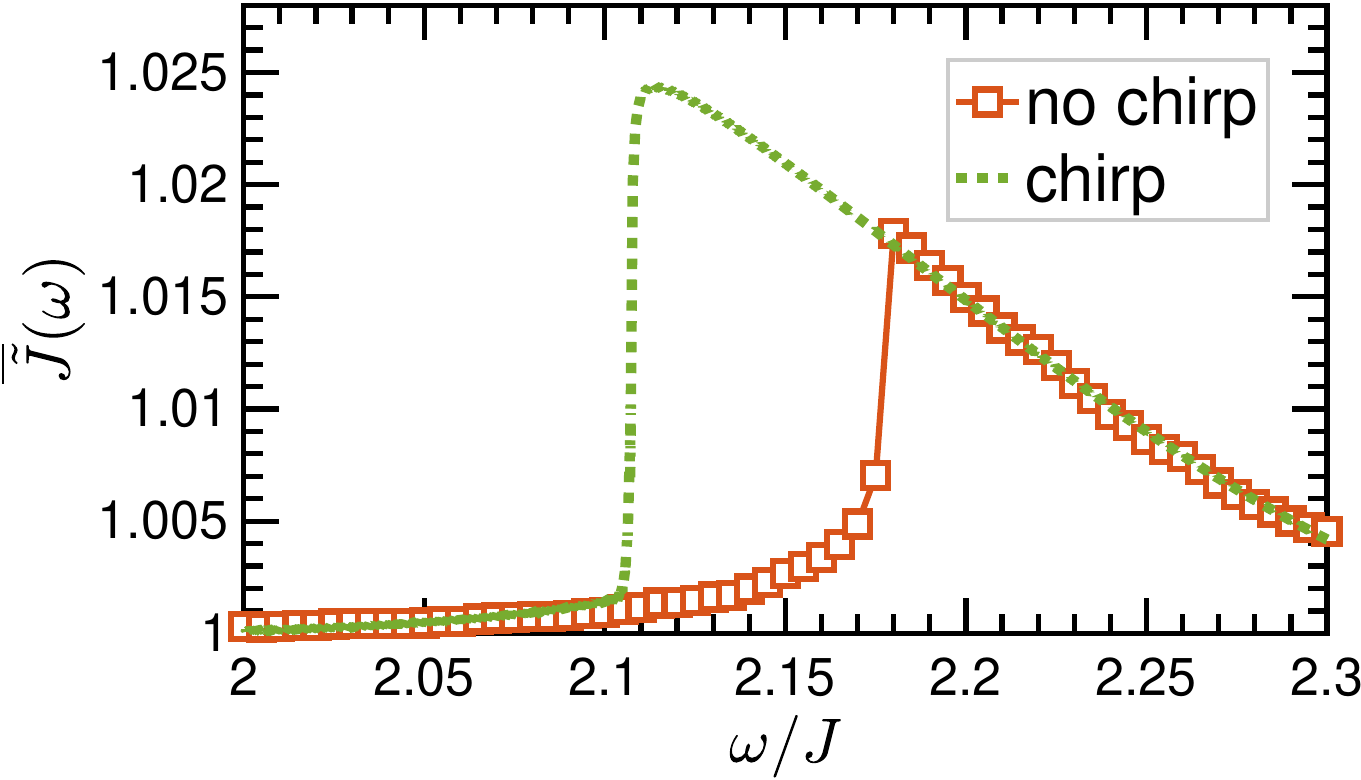}
\setlength\abovecaptionskip{1pt}
\caption{\textbf{Chirp protocol for tuning enhancement of magnetism and first-order phase transition in SmFeO$_3$.} Linearly chirped isotropic magnetic interaction with $\mathcal{E}(t)$ from Eq.~\eqref{eq:chirp} for \{$\omega_1/J = 2.3$, $\tau_1 =500$ ps\} and \{$\omega_2/J = 2$, $\tau_2 =5000$ ps\}. Using this chirp allows one to nearly double the size of $\tilde{J}$ compared to the unchirped case before a second transition occurs at $\omega/J = 2.1$. Here, $\omega_0 = 2.5 J$, $\Delta = 1.2$, $\mathcal{A}_0/\gamma_{\rm ph} = 0.1$, $g_{\rm q}/J = 0$, $g_{\rm l}/J = 0.5$, $\gamma_{\rm ph}/\omega_0 = 0.05$, and $\gamma_{\rm s}/J = 0.01$.} 
\label{f3}
\end{figure}

Having uncovered the phase transition physics in our model, we now ask whether it can be used to engineer the size of the interaction strength. Inspired by our recent work on electronic materials~\cite{yarmohammadi2023nonequilibrium}, we consider a
linear chirp protocol given by\setlength{\abovedisplayskip}{3pt}	\setlength{\belowdisplayskip}{1pt}	\begin{equation} \label{eq:chirp}
\mathcal{E}^{\rm chirp}(t) = \mathcal{A}_0 \cos(\omega_1\,t + \left[\frac{(\omega_2-\omega_1)(t-\tau_1)}{\tau_2-\tau_1}\right]t)\, ,
\end{equation}where $\omega_1/J = 2.3$, $\tau_1 =500$ ps, $\omega_2/J = 2$, and $\tau_2 =5000$ ps. In this protocol, we slowly ramp the laser frequency down from $\omega_1$ to $\omega_2$ by passing through the original phase transition point; in doing so, stronger magnetic interactions~(nearly twice the enhancement compared to non-chirped interactions) can be generated through a hysteresis loop of adiabatic NESSs, as shown in Fig.~\ref{f3}. Accordingly, it is also accompanied by another first-order phase transition at $\omega/J = 2.1$. 

\textit{Experiment.}---Continuous driving unavoidably leads to heating of the sample, and remediation of thermal effects is necessary~\cite{yarmohammadi2020dynamical,yarmohammadi2023nonequilibrium}. To provide a measurable quantity in the experiment, we roughly estimate the timescale for reaching the NESS and phase transition. To this end, we focus on the dumped power, i.e., the part of input laser power that flows directly from the driven phonon to the bath, given by $\mathcal{P}_{\rm du}(t) = w a \rho\, \omega_0 \gamma_{\rm ph} n_{\rm ph}(t)$. In doing so, we choose SmFeO$_3$ sample with a thickness $w = 20$ nm, area $a = 2$ mm$^2$, and molar density $\rho \simeq 0.028$ mol.cm$^{-3}$~\cite{Villars2016_sm_isp_sd_1831610}. We assume that the sample is in touch with a metal block~(heat sink) to take up the thermal energy; the sink should be at a low-temperature $T \simeq 2-5$ K with a mass $m_{\rm b} \simeq 2-5$ g and specific heat $C(T) \simeq 1-3\,\times 10^{-4}\, T\,$JK$^{-2}$g$^{-1}$ such that overheating of the block at the phase transition point with $\overline{n_{\rm ph}}^{\rm iso} \simeq 0.01$, occurs at:\setlength{\abovedisplayskip}{3pt}	\setlength{\belowdisplayskip}{1pt}	\begin{equation}\label{eq_7}
\,t^{\rm heat} = \frac{m_{\rm b} \int_0^T C(T) \, d\,T}{\overline{\mathcal{P}_{\rm du}}^{\rm iso}} \simeq 50 \, {\rm ns} ,
\end{equation}
for the isotropic scenario. This timescale to reach the NESS is much larger than the obtained ps timescale in our simulation. Thus, the computed long-time behavior is readily observable if we have such a heat sink to dampen thermal effects. This, in turn, means that the uncovered phase transition can be experimentally accessible under these conditions. 


\textit{Conclusions.}---In contrast to studies that have relied on equilibrium treatments and static approaches to tune the magnetic responses of a system, we propose the dynamical magnetophononic mechanism -- modulation of magnetism through phonon -- beyond equilibrium. The rare-earth orthoferrite SmFeO$_3$, as a functional material in both science and industry, yields potentially unknown features due to its multiple magnetic orderings associated with an unusual nonlinear spin-phonon coupling, however, its inherent magnetic exchange couplings are weak for spintronic applications. For the dynamic control of the magnetic response of SmFeO$_3$, we apply a continuous laser field. We consider dissipation effects when pumping energy into the system to achieve nonequilibrium steady states at long times. For the on-resonance drive of the phonon, inherently weak Sm-Fe magnetic interactions are enhanced~(manipulation of the magnonic band), which is relevant for spintronic applications. 

In connection to experimental setups; we scan the system's responses to the pump field when the laser and phonon are chosen to be in the off-resonance regime, which unveils an intriguing phenomenon out of equilibrium, referred to as dynamical magnetophononic first-order phase transition. In contrast to the electron-phonon coupled chains~\cite{Sous2021,Kennes2017,yarmohammadi2023nonequilibrium} in which the quadratic model reacts more strongly to a pump than the linear model, linear spin-phonon coupling is sufficient to observe this phenomenon and is, indeed, more effective than quadratic coupling in enabling phase transitions and engineering magnetic interactions. We finally propose a chirp drive protocol to engineer both the first-order phase transition and the enhancement rate of magnetism. 

\section*{Acknowledgments}
This work was performed with support from the National Science Foundation (NSF) through award numbers MPS-2228725 and DMR-1945529 and the Welch Foundation through award number AT-2036-20200401 (MK and MY). Part of this work was performed at the Aspen Center for Physics, which is supported by NSF grant No. PHY-1607611, and at the Kavli Institute for Theoretical Physics, which is supported by NSF grant No. NSF PHY-1748958. This project was funded by The University of Texas at Dallas Office of Research and Innovation through the SPIRe program. 
}
	\bibliography{bibliography}
	
\end{document}